# Metal-organic Pulsed Laser Deposition for Complex Oxide Heterostructures

*Jung-Woo Lee,*[1,11] *Jieun Kim,*[1,11] *Anthony L. Edgeton,*[1,11] *Tula R. Paudel,*[2,10] *Neil Campbell,*[3] *Brenton A. Noesges,*[4] *Jonathon L. Schad,*[1] *Jiangfeng Yang,*[1] *Katelyn Wada,*[5] *Jonathan Moreno-Ramirez,*[5,6] *Nicholas Parker,*[5] *Yulin Gan,*[7] *Hyungwoo Lee,*[1] *Dennis V. Christensen,*[7] *Kitae Eom,*[1] *Jong-Hoon Kang,*[1] *Yunzhong Chen,*[7] *Thomas Tybell,*[8] *Nini Pryds,*[7] *Dmitri A. Tenne,*[5] *Leonard J. Brillson,*[4,9] *Mark S. Rzchowski,*[3] *Evgeny Y. Tsymbal,*[2] *and Chang-Beom Eom*[1,12]*

[1]Department of Materials Science and Engineering, University of Wisconsin-Madison, Madison, WI 53706, USA.
[2]Department of Physics and Astronomy, Nebraska Center for Materials and Nanoscience, University of Nebraska, Lincoln, NE 68588, USA.
[3]Department of Physics, University of Wisconsin-Madison, Madison, WI 53706, USA.
[4]Department of Physics, The Ohio State University, Columbus, OH 43210, USA.
[5]Department of Physics, Boise State University, Boise, ID 83725, USA.
[6]Riverstone International School, Boise, ID 83716, USA.
[7]Department of Energy Conversion and Storage, Technical University of Denmark (DTU), DK-2800 Kongens Lyngby, Denmark.
[8]Department of Electronic Systems, Norwegian University of Science and Technology, 7491 Trondheim, Norway.
[9]Department of Electrical and Computer Engineering, The Ohio State University, Columbus, OH 43210, USA.
[10]Present address: Department of Physics, South Dakota School of Mines and Technology, Rapid City, South Dakota 57701, USA.
[11]These authors contributed equally.
[12]Lead contact
*Correspondence: eom@engr.wisc.edu

## SUMMARY

Point defects in complex oxide thin films play a critical role in determining material properties but remain challenging to control with precision. This study introduces metal-organic pulsed laser deposition (MOPLD) as a novel synthesis technique for the precise manipulation of these defects, using $LaAlO_3/SrTiO_3$ (LAO/STO) as a model system. By employing titanium tetraisopropoxide (TTIP) as the titanium precursor, MOPLD achieves refined stoichiometric control in STO layers while preserving their structural integrity, as confirmed by X-ray diffraction and Raman spectroscopy. Depth-resolved cathodoluminescence spectroscopy and density functional theory calculations reveal that increasing TTIP flux during STO growth enhances the $[Ti_{Sr}]/[V_{Sr}]$ ratio and reduces the $[V_O]$ concentration. These defect modifications lead to a significant improvement in the low-temperature mobility of the two-dimensional electron gas at the LAO/STO interface, evidenced by distinct Shubnikov–de Haas oscillations. This work underscores the potential of MOPLD to advance defect engineering in complex oxide heterostructures, opening new avenues for quantum material research.

## INTRODUCTION

Point defects are ubiquitous in any solid material. It is impossible to obtain *perfect* solid materials because thermodynamics always enforce certain amounts of point defects due to the entropy of mixing.[1] Even though the amount of point defects is small with respect to host atoms, they play a major role and often control the functionalities of materials,[2] such as ferroelectricity[3] and metal-insulator transitions.[4] Accordingly, there have been many reports on point defects and their effects on material properties.[2,5] In complex oxide thin films and their heterostructures, however, the ability to precisely control point defects is still a challenging subject.[6] In a variety of deposition techniques including sputtering,[7] molecular beam epitaxy (MBE),[8] and pulsed laser deposition (PLD),[9] coupled with *in-situ* control of elements such as from quartz crystal microbalance or reflection high-energy electron diffraction (RHEED), the ability to control elements and stoichiometry is still limited by the accuracy of controllable flux which is on the order of ~1%,[10] corresponding to defect concentrations of ~$10^{20}$ cm$^{-3}$.[11] Such defect levels typically lead to large defect scattering and low electron mobility incompatible with observing quantum phenomena in most materials systems.[12] The ability to control the cation stoichiometry with the accuracy less than 1% would not only allow for accurate matching of the amount of constituent cations, but also opens up a new way to engineer point defects and thereby to enhance specific properties governed by thermodynamic relationships between stoichiometry and defect formations.

Controlling cation stoichiometry with such high accuracy requires a near-equilibrium process, where stoichiometry is determined by thermodynamics rather than flux ratios. For example, a near-equilibrium hybrid MBE process has successfully produced stoichiometric SrTiO$_3$ (STO) films with electron mobility exceeding that of bulk single crystal.[13] Building on this, developing a PLD-based near-equilibrium process could further expand its applicability, enabling more precise control over cation stoichiometry and point defect engineering in complex oxide thin films and heterostructures. Such advancements could also facilitate complete control of individual point defects by leveraging the thermodynamic relationships between stoichiometry and defect formation.

Strontium titanate (STO) is one of the most extensively studied perovskites, known for its diverse ground-state physical phenomena and widespread use as a template for epitaxial growth.[14] In this work, we introduce a metal-organic pulsed laser deposition (MOPLD) technique that enables precise control of point defects in STO thin films. This approach establishes a reliable process window for producing stoichiometric STO films while preserving their structural integrity. Using the MOPLD method, we acheived STO films with high [antisite Ti (Ti$_{Sr}$)]/[strontium vacancy (V$_{Sr}$)] ratio, effectively reducing the equilibrium oxygen vacancy (V$_O$) concentration. This defect optimization led to enhanced electron mobility in LaAlO$_3$/SrTiO$_3$(LAO/STO) heterostructures, evidenced by clear Shubnikov–de Haas oscillations. This targeted defect management enables the fabrication of high-quality heterostructures and interfaces, opening new avenues for high-mobility applications in complex oxide-based quantum devices.

## RESULTS AND DISCUSSION

### Defect Engineering in STO thin films via MOPLD

In the MOPLD process, titanium tetraisopropoxide (TTIP) serves as the titanium source, introduced concurrently with the laser ablation of a strontium oxide (SrO) target, enabling precise control over the deposition and stoichiometry of the resulting thin film. Comprehensive details regarding the MOPLD apparatus including its configuration is provided in Figure S1. Figure 1A and 1B illustrate the MOPLD process for forming SrO and TiO$_2$ layers, along with the associated defect structures. We propose that MOPLD growth mechanism for STO involves selective adsorption and decomposition of TTIP on a SrO-terminated surface, where it forms STO, while it desorbs from TiO$_2$-terminated surface (Figure 1A, see supplemental information (Figure S2) for details).[15] Although STO is typically considered a line compound,[16] bulk studies at elevated temperature (~1000°C) indicate a limited solubility (<0.1%) between STO and TiO$_2$.[17] Through MOPLD, the Ti/Sr cation ratio can be finely tuned within this narrow solubility range by adjusting the TTIP flux (Figure S3). This precise control of cation stoichiometry enables the adjustment of defect concentrations to levels below ~$10^{19}$ cm$^{-3}$, leveraging the thermodynamic relationship between stoichiometry and defect formation.

In a slightly Ti-rich STO within the solubility limit, the formation of cation point defects $V_{Sr}$ and $Ti_{Sr}$ are energetically favorable.[18] Based on the MOPLD growth mechanism, TTIP can only react with a SrO monolayer where Schottky pairs ($V_{Sr}^{2-}$–$V_O^{2+}$) exist.[19] When additional TTIP is supplied, $Ti^{4+}$ ions occupy $V_{Sr}^{2-}$ sites, forming $Ti_{Sr}^{2+}$ defects (Figure 1B). The [$Ti_{Sr}^{2+}$]/[$V_{Sr}^{2-}$] ratio plays a crucial role, as it affects the equilibrium amount of $V_O$. Since $V_O$ is positively charged,[20] increasing the [$Ti_{Sr}^{2+}$]/[$V_{Sr}^{2-}$] ratio reduces the $V_O$ concentration in accordance with charge neutrality. The defect chemistry can be represented by the equation (1):

$$2V_{Sr}^{2-} + 2V_O^{2+} + TiO_2 = Ti_{Sr}^{2+} + V_{Sr}^{2-} + 2O_O^x \quad (1)$$

This selective defect formation driven by excess TTIP is supported by defect formation energy calculation by density functional theory (DFT). In Figure 1C, the formation energies of each defect are plotted as a function of $\Delta\mu_{Ti}$, where is $\Delta\mu_{Ti}$ is related to $\Delta\mu_O$ by $\Delta\mu_{Ti} = 6\Delta\mu_O$, reflecting the chemical potential shifts from stoichiometric STO to $TiO_2$ (see supplemental information (Figure S4) for details). The results indicate that increasing TTIP flux raises the formation energy of $V_O^{2+}$ while reducing the formation energy difference between $Ti_{Sr}^{2+}$ and $V_{Sr}^{2-}$. This enables STO films to achieve a higher [$Ti_{Sr}^{2+}$]/[$V_{Sr}^{2-}$] ratio with reduced [$V_O^{2+}$]. Minimizing $V_O$ in STO is essential, as $V_O$ is a known dominant scattering center at low temperatures.[21,22] Differences in scattering mechanisms are illustrated by calculating the density of conducting states associated with each defect. The localized nature of defect states from $Ti_{Sr}$ (~ 2 unit cells, Figure 1D) contrasts with the delocalized states generated by $V_O$ (Figure 1E), underscoring the higher scattering impact of $V_O$ compared to $Ti_{Sr}$.

## Structural stoichiometric growth window in STO thin films via MOPLD

Ideally controlling the defect population within the solubility limit is critical for robustness, as any excess Sr or Ti beyond ~1% leads to volume expansion, allowing for a higher concentration of $V_O$.[23] To investigate this, we homoepitaxially grew 25 u.c. of STO thin films on (001) STO single crystal substrates using the MOPLD method (Figure S5). Figure 2A shows out-of-plane *θ-2θ* XRD patterns of STO films as a function of TTIP gas inlet pressure. Since the STO lattice parameter is highly sensitive to stoichiometry, XRD reflections observed away from the single crystal substrate peak indicate cation off-stoichiometry.[23] The results reveal a stoichiometric STO growth window over a wide range of TTIP gas inlet pressure, which serve as an indicator of TTIP flux during the growth (11−21 mTorr), where the peaks of STO films perfectly align with those of the STO single crystal substrates. At slightly lower TTIP gas inlet pressure (9 and 10 mTorr), a secondary phase corresponding to $Sr_2TiO_4$ and off-stoichiometric STO is detected. At higher TTIP gas inlet pressure (22 and 23 mTorr), shoulders at lower angle relative to the ideal STO (002) peak suggest the formation of Ti-rich STO films.

For further characterization of STO structures within the identified growth window, low temperature Raman spectroscopy was performed. Ideal STO has a cubic perovskite structure (space group Pm-3m) with 12 optical phonon modes.[24] Due to their odd symmetry with respect to the inversion center, 1$^{st}$ order Raman peaks are absent in a stoichiometric structure. However, cation off-stoichiometry, whether Sr- or Ti-rich, breaks the inversion symmetry of STO, generating the 1$^{st}$ order peaks in Raman spectra, such as the $LO_3$ or $TO_4$ modes.[24] Therefore, low-temperature Raman spectroscopy serves as a sensitive probe for detecting non-stoichiometry resulting from synthesis conditions. The Raman spectra of STO films measured at 10 K are shown in Figure 2B. For Ti-rich STO films grown at TTIP gas inlet pressures of 22 and 23 mTorr, the $LO_3$ and $TO_4$ peak intensities are stronger compared to STO films grown at TTIP gas inlet pressure between 11 and 21 mTorr, consistent with the stoichiometric growth process window determined by XRD. To quantiatively assess stoichiometry, we use the normalized intensity $I_{TO4}/I_{2nd\ order}$ as shown in Figure 2C. Within the TTIP gas inlet pressure range of 12–21 mTorr, the $I_{TO4}/I_{2nd\ order}$ values are significantly lower than those of films grown at higher pressures, defining this range as the *structural* stoichiometric growth window of STO. Notably, the Raman-determined growth window is slightly narrower than that observed by XRD. For instance, the STO film grown at 11 mTorr of TTIP gas inlet pressure exhibits higher 1st order peak intensities compared to those grown within the 12–21 mTorr pressure range, indicating slight non-stoichiometry.

**Reduced defect scattering and enhanced mobility at LAO/MOPLD-STO Interfaces**

We use the electron mobility at low temperature (< 5 K) as a figure of merit to access *cleanness* of the system, where impurity scattering dominantes.[25] Specifically, we measure the two-dimensional electron gas (2DEG) at the LAO/STO interface to correlate cation stoichiometry control discussed earlier with the transport properties (Figure S6). For this purpose, 5 u.c. LAO thin films were prepared by conventional PLD on top of 12 u.c. STO films grown by MOPLD.

The LAO/MOPLD-STO interface exhibits higher electron mobility at 2 K compared to LAO deposited on commercially available STO substrate (Figure 3A), indicating fewer dominant scattering centers, such as oxygen vacancies, consistent with DFT calculations (Figure 1E). Since electron mobility also depends on carrier density, we investigated the relationship between electron mobility and carrier concentration measured at 2 K. As shown in Figure 3A–C, the LAO/MOPLD-STO interface (labeled "Film") consistently exhibits higher mobility and residual resistivity ratio (RRR) than the bulk single crystal-based heterostructure (labeled "Bulk") across all carrier concentrations. Electron mobility and carrier concentrations were extracted from linear Hall curves (Figure 3D), confirming carrier transport in a single conduction band. Another signature of high sample quality and low defect density is the observation of well-defined Shubnikov-de Haas (SdH) oscillations at low temperatures and high magnetic fields (Figure 3E), where the longitudinal resistance $R_{xx}$ oscillates as the perpendicular magnetic field $B$ is varied. From the SdH oscillations, we extract an effective mass of 1.0 $m_e$ ($m_e$ is the bare electron mass, Figure S7), which is lower than the values reported LAO/single-crystal STO interfaces grown under optimized conditions[26], while the partial carrier density remains 1.7×10$^{12}$ cm$^{-2}$ (Figure S8).

**Depth-Resolved defect analysis in MOPLD-Grown STO**

To investigate the point defect population in MOPLD-grown STO and its correlation with low-temperature mobility, we performed depth-resolved cathodoluminescence spectroscopy (DRCLS).[27,28] Monte Carlo simulations were used to optimize the incident beam voltage, ensuring effective probing of both the MOPLD-grown STO layer and the bulk STO region (Figure S9). Figure 4A presents DRCLS spectra of the LAO/STO sample, where the STO film was grown at the TTIP inlet pressure of 21 mTorr. Probe beam voltages of 0.3 kV and 2.5 kV were applied to selectively probe the STO film and bulk regions, respectively. A clear intensity difference is observed between the two regions in the energy range of ~1.5 to ~3.4 eV when normalized to the band gap energy (3.6 eV). Figure 4b shows the relative point defect ratio [Ti$_{Sr}$]/[V$_{Sr}$] and the relative concentration of V$_O$ for the STO film and bulk regions. In MOPLD-grown STO, the [Ti$_{Sr}$]/[V$_{Sr}$] ratio is higher than in bulk STO, while the concentration of V$_O$ is reduced in the thin film. This defect interplay contributes to the enhanced mobility in LAO/STO interfaces, consistent with the previous finding on defect-engineered LAO/STO systems with manganite buffer layers.[22,29]

**Conclusions**

We have successfully developed the MOPLD synthesis approach to precisely manipulate individual point defects in complex oxide systems. By combining spectroscopic, crystallographic, and electrical property measurements, we have established the fundamental mechanism of MOPLD and its role in point defect control, leveraging the thermodynamic relationships between stoichiometry and defect formation. Specifically, the MOPLD process enables precise control of the cation stoichiometry of STO to within a solubility limit below 0.1%, across a relatively wide structural stoichiometry growth window. This precise defect control has significant implications for the macroscopic properties of complex oxides, potentially uncovering new physical phenomena, including quantum effects that have previously been obscured by point defects. For instance, achieving high mobility p-type conductivity in oxide materials remains a challenge due to the formation of oxygen vacancies.[30] The MOPLD approach offers a pathway to overcome such challenges, opening new opportunities to explore the physics and fundamental properties of complex oxides.

## METHODS

### Sample fabrications

Single crystalline STO (001) substrates were used for this study. STO substrates are soaked in DI-water for 30 min and in buffered-BHF for 1 min to make $TiO_2$- terminated surface. By subsequent annealing at 1000°C for 6 h in 1 atm $O_2$ atmosphere, we could obtain an atomically smooth surface. In the MOPLD process, TTIP gas served as a working gas during laser ablation of the SrO single crystal target (Figure S1, Supporting Information). A KrF excimer laser ($\lambda$=248 nm) was used with ~0.7 J/cm$^2$ of energy fluence and 2 Hz of repetition rate on the SrO target. TTIP gas was supplied by a heated pyrolytic boron nitride (PBN) showerhead gas injector and a PID controlled variable leak valves controlled the TTIP gas inlet pressure using closed loop feedback by a capacitance manometer vacuum gauge. During the STO film growth, the substrate temperature was kept at 900 °C. The sample thickness was evaluated from RHEED oscillations (Figure S5, Supporting Information). Following the growth of STO film, samples were cooled down to 600 °C and then oxygen gas was backfilled to ~600 Torr for post-annealing. For LAO layers, a conventional PLD method was employed. The growth temperature, oxygen partial pressure, energy fluence and repetition rate for LAO are 750 °C, 7.5×10$^{-5}$ Torr, ~1.0 J/cm$^2$, and 1 Hz, respectively. After growth, samples were annealed at 600 °C in ~150 Torr of $O_2$ for 1 h in order to achieve an equilibrium oxygen concentration.

### Theoretical calculations

For the calculation of defects, we use density functional theory (DFT) band structure approach as implemented in Vienna *ab initio* simulation package (VASP).[31] The projected augmented wave (PAW) method is used to approximate the electron-ion potential.[32] To treat exchange and correlation effects we use both the local density approximation (LDA)[33] and the semi-empirical LDA+U method[34] within a rotationally invariant formalism[35], for a better description of the localized transition metal *d* electrons. Here we choose (*U-J*) = 5 eV for the 3*d* orbitals of Ti atoms, as this value of *U* provides good description of the lattice parameters; calculated lattice constant is 3.9 Å similar to that in experiment and improve band gap to 2.4 eV and include most of defects states in the band gap as suggested by higher order functional such as HSE. Defect calculations were performed in 135 atoms $3 \times 3 \times 3$ cubic supercell. After creating one such defect in the perfect supercell, we relax internal coordinate until the Hellman-Feynman forces are less than 0.01 eV/Å. In the calculation, we use a kinetic energy cutoff of 340 eV as kinetic energy cutoff and $6 \times 6 \times 6$ Monkhorst-Pack grid[36] of *k* points Brillouin zone integration. In all calculations, we turn on the spin polarization to include the effect of the local moment introduced by defect. To establish an ionized (charged) defect, we add or remove electrons to the system and include a compensating jellium background. Formation enthalpy of the defect *D* is energy cost to add (remove) an atom of charge *q* to (from) otherwise a perfect host and is calculated using relation: $\Delta H_f(D,q) = E(D,q) - E_H + \mu_{removed} - \mu_{added} + q(E_V + E_F)$, where E(*D*,*q*) is the energy of the host with the defect, $E_H$ is energy without defect, and $E_F$ is the electrochemical potential of the charge *q* that is usually measured with respect to the host valence band maximum ($E_V$).

### Structural Analysis

The crystal structure of samples was analyzed using a high-resolution four-circle XRD machine (Bruker D8 advance) using CuK$_{\alpha 1}$ radiation. Raman spectra were measured using a Horiba Jobin Yvon T64000 triple spectrometer equipped with a liquid-nitrogen-cooled multichannel charge-coupled device detector. Spectra were recorded in backscattering geometry in the temperature range of 10–300 K, using a variable temperature closed cycle He cryostat. The 325 nm line of a He–Cd laser line was used for excitation; laser power density was below 0.5 W/mm$^2$ at the sample surface, low enough to avoid any noticeable local heating. The TO$_4$ peak (at ~550 cm$^{-1}$) was used for analysis of symmetry breakdown due to point defects, since this peak is the most distinctive and does not overlap with the second order features. The ratio of the integrated intensity of the TO$_4$ peak to that of the 2$^{nd}$ order peak at 620 cm$^{-1}$ (plotted in Figure 2C) was used to determine the structural growth window.

### Electrical Measurement

The SdH measurements were performed using a four-probe Van der Pauw method with ultrasonically wire-bonded aluminium wires as electrodes. (Figure S6a, Supporting Information) A CRYOGENIC cryogen-free measurement system was employed to characterize the temperature-dependent SdH oscillations in perpendicular magnetic field up to 16 T at low temperatures from 1.8-3.1 K in steps of 0.2 K. The amplitude of the quantum oscillations at different temperatures was extracted from these measurements by subtracting a 2$^{nd}$ order polynomial background. The data analysis is based on the method published elsewhere.[25,37,38]

### Depth-Resolved Cathodoluminescence Spectroscopy

Monte Carlo simulations of electrons in solids (CASINO) is used for simulated electron energy loss in solids, which determines depth to which electron beam reaches (Figure S10, Supporting Information). For the analysis of 12 u.c. MOPLD-STO layer underneath 5 u.c. LAO layer, the beam voltage of 0.3 kV is used. The sample surface was grounded with fine copper grid to reduce surface charge buildup. For 0.3 kV low beam voltage, 2 mA constant emission current (variable power), average of multiple 20s acquisitions (up to 100), e-beam deflected off-sample which is "dark counts" is considered as chamber background signal. The beam deflected onto sample which is "bright counts" is evaluated as CL signal + background signal. All scans are taken at room temperature. A ZnO crystal of roughly identical thickness to the sample was used to optimize beam focus and align the beam spot with collection optics for each beam voltage.

### RESOURCE AVAILABILITY

#### *Lead contact*
Further information and requests for resources and reagents should be directed to and will be fulfilled by the lead contact, Chang-Beom Eom (eom@engr.wisc.edu).

#### *Materials availability*
This study did not generate new unique reagents.

#### *Data and code availability*
All experimental data are available upon reasonable request to the lead contact.

### ACKNOWLEDGMENTS

C.B.E. acknowledges support for this research through the Gordon and Betty Moore Foundation's EPiQS Initiative, Grant GBMF9065 and a Vannevar Bush Faculty Fellowship (ONR N00014-20-1-2844). Transport measurement at the University of Wisconsin–Madison was supported by the US Department of Energy (DOE), Office of Science, Office of Basic Energy Sciences (BES), under award number DE-FG02-06ER46327 (C.B.E.). N.P. and D.V.C. acknowledge the support of Novo Nordisk Foundation Challenge Pro-gramme 2021: Smart nanomaterials for applications in life-science, BIOMAG Grant NNF21OC0066526. NP acknowledge the support from the ERC Advanced "NEXUS" Grant 101054572 and the Danish Council for Independent Research Technology and Production Sciences for the DFF- Research Project 3 (grant No 00069B). Raman measurements at Boise State University were supported by the National Science Foundation Grant DMR-2104918, and by M.J. Murdock Charitable Trust "Partners in Science" program.

**AUTHOR CONTRIBUTIONS**

J.W.L. and C.B.E. conceived the project. C.B.E., E.Y.T., M.S.R., L.J.B., D.A.T., N.P., and T.T. supervised the experiments. J.W.L., J.K. A.L.E., J.L.S., J.Y., and C.B.E. built the MOPLD system, and fabricated and characterized thin-film samples. T.R.P. and E.Y.T. performed theoretical calculations. N.C., Y.G., H.L., D.V.C., Y.C., N.P., and M.S.R. conducted electrical transport measurements. B.A.N. and L.J.B. performed depth-resolved cathodoluminescence spectroscopy. K.W., J.M.-R., N.P., and D.A.T. carried out temperature-variable Raman spectroscopy. K.E. and J.H.K. contributed to modeling the MOPLD process and defect generation. J.W.L., J.K. A.L.E., T.R.P., N.C., B.A.N., J.L.S., Y.G., D.V.C., K.E., Y.C., T.T., N.P., D.A.T., L.J.B., M.S.R., E.Y.T., and C.B.E. prepared the manuscript. C.B.E. directed the overall research.

**DECLARATION OF INTERESTS**

The authors declare no competing interests.

**SUPPLEMENTAL INFORMATION**

Supplemental information is prepared.

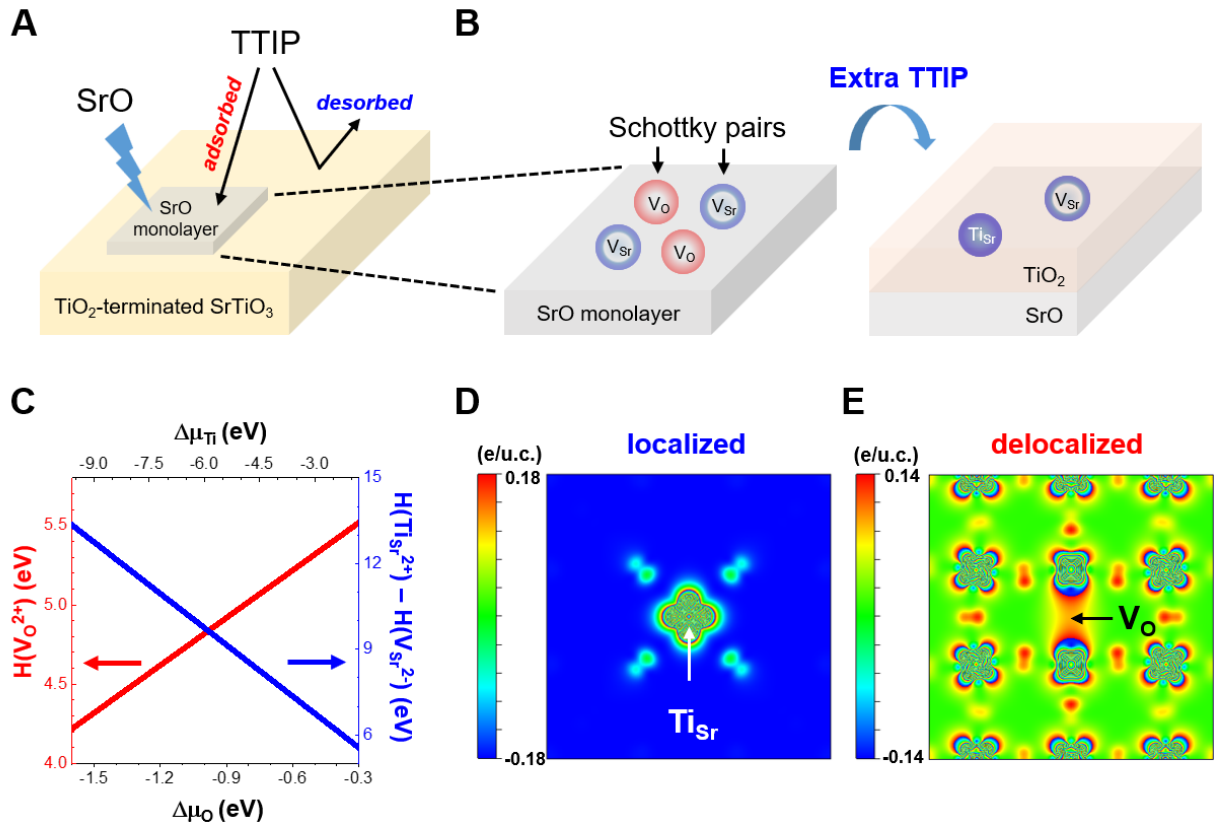

**Figure 1. Schematic of STO film growth and defect formation via MOPLD.**

(A) Schematic illustration of the MOPLD process for STO growth, where SrO is suppllied by laser ablation and TTIP is introduced through a gas injector.

(B) Schematic of vacancy formation and annihilation. Schottky-type defects in the SrO monolayer can be annihilated by additional $TiO_2$ supplied through TTIP decomposition.

(C) Defect formation energy of $V_O^{2+}$ and the energy difference between $Ti_{Sr}^{2+}$ and $V_{Sr}^{2-}$ as a function of the chemical potential change of O and Ti. Note that $\Delta\mu_O$ and $\Delta\mu_{Ti}$ are coupled to account for the chemical potential change of $TiO_2$ in STO (see supplemental Information for details).

(D and E) Charge density distribution in STO matrix with the introduction of $Ti_{Sr}$ (D) and $V_O$ (E), respectively, under n-type condition.

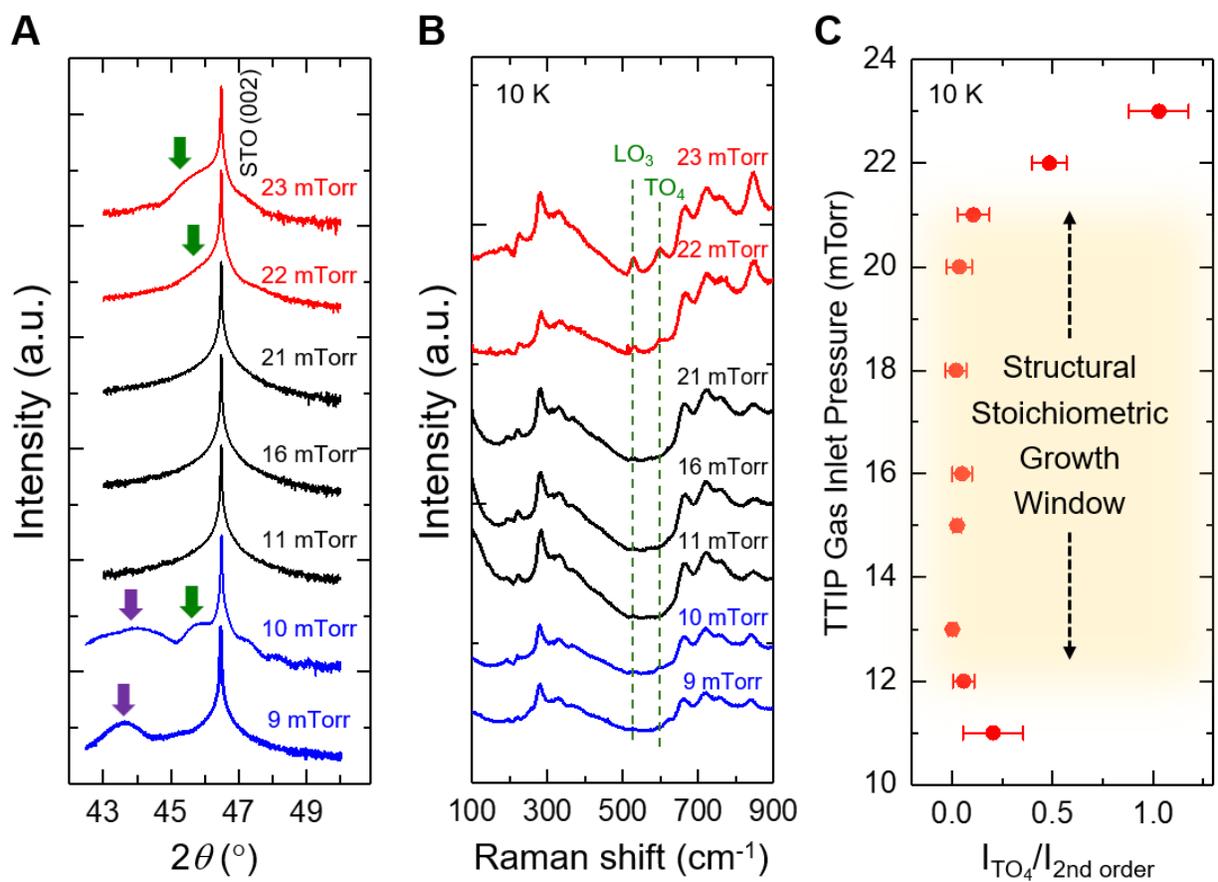

**Figure 2. Structural analysis of STO films grown by MOPLD.**

(A) Out-of-plane $\theta$-$2\theta$ XRD patterns of STO films grown on (001) STO substrates. The (002) peak positions of off-stoichiometric STO films and $Sr_2TiO_4$ are indicated by the green and purple arrows, respectively.

(B) Raman spectra of STO films on (001) STO substrates measured at 10 K. Data for Sr-rich, stoichiometric, Ti-rich STO films (as determined by XRD) are shown in blue, black, red, respectively.

(C) Normalized $TO_4$ intensity in Raman spectra (B) as a function of TTIP inlet pressure. Sample with the $Sr_2TiO_4$ phase (grown at 9 and 10 mTorr) are excluded. The yellow area indicates the structural stoichiometric growth window determined from Raman spectra at 10 K.

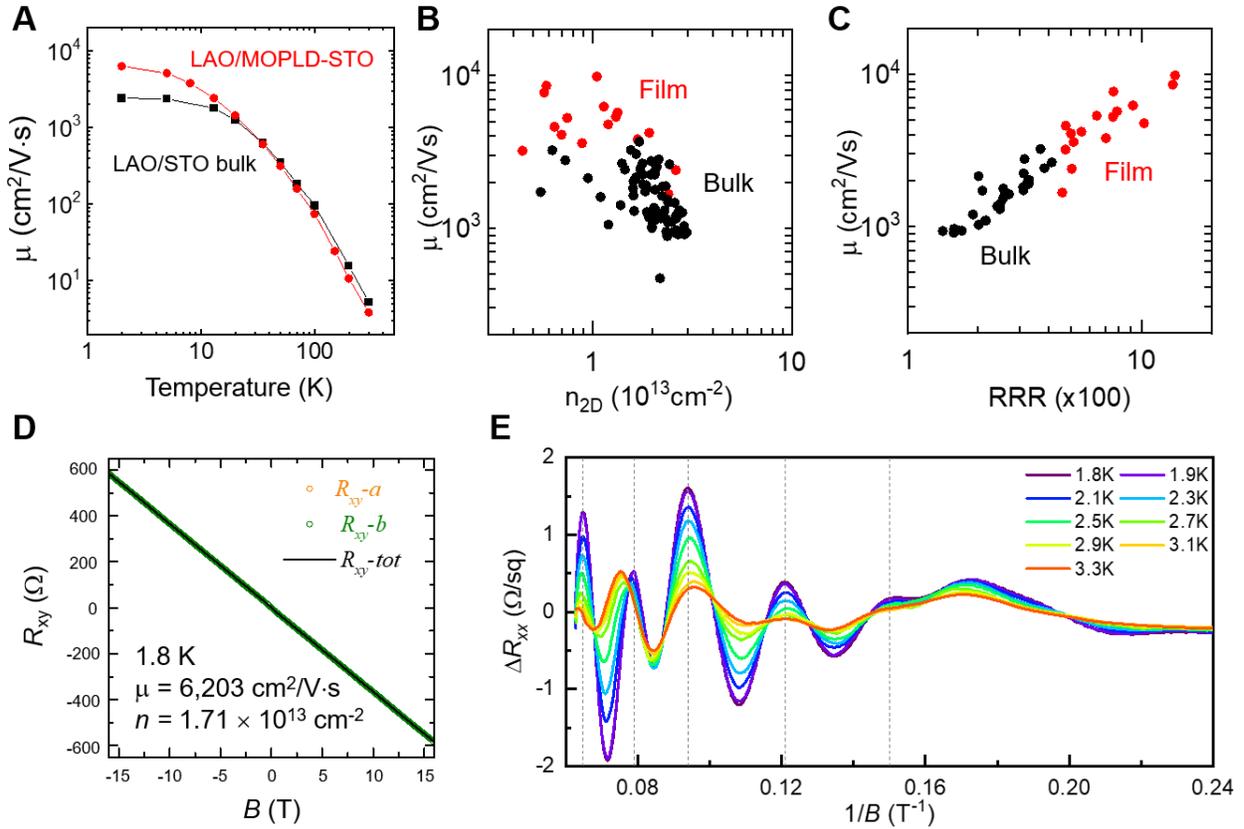

**Figure 3. Electrical properties of LaAlO$_3$/SrTiO$_3$ interfaces.**

(A) Electron mobility at 2 K in LAO/STO bulk and LAO/MOPLD-STO film.

(B) Distribution of electron mobility ($\mu$) and 2D carrier density ($n_{2D}$) extracted from Hall measurements at $T = 2$ K.

(C) Distribution of electron mobility ($\mu$) and residual resistivity ratio (RRR) from Hall measurements at $T = 2$ K.

(D) Hall resistance ($R_{xy}$) as a function of magnetic field at 1.8 K for LAO/MOPLD-STO sample.

(E) Longitudinal resistance ($R_{xx}$) versus inverse magnetic field ($1/B$) after background subtraction, showing Shubnikov–de Haas oscillations in the temperature range 1.8–3.3 K.

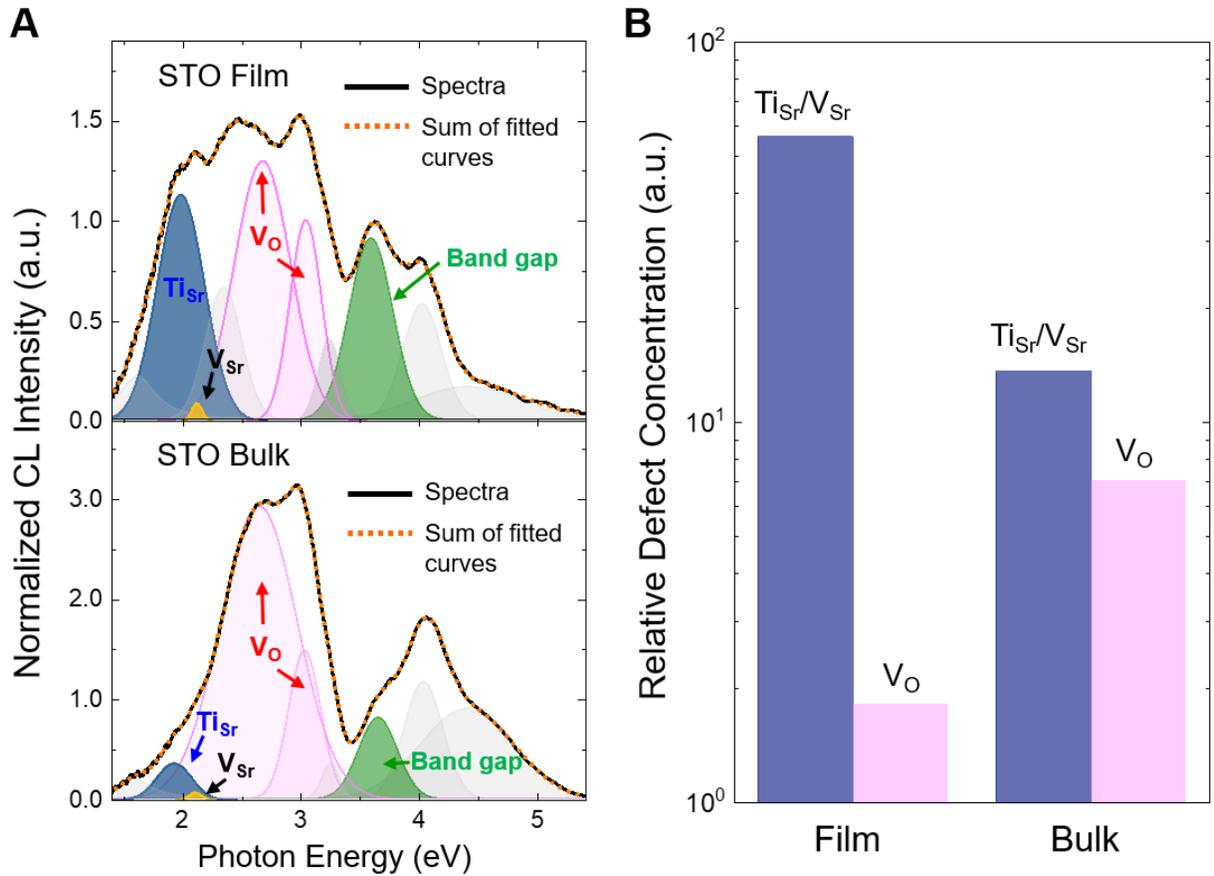

**Figure 4. Depth-resolved cathodoluminescence spectroscopy (DRCLS) measurement.**

(A) DRCLS spectra of a 5 u.c. LAO/12 u.c. MOPLD-grown STO film on a (001) STO substrate, showing data for the STO film and bulk regions.

(B) Relative defect concentration in the STO film and bulk. The relative oxygen vacancy concentration was determined by summing the $V_O^{2+}$ and $V_O$-related peak areas, normalized to the bandgap area near 3.6 eV. The relative ratio of $[Ti_{Sr}]/[V_{Sr}]$ was evaluated as the area ratio of $Ti_{Sr}^{2+}$ to $V_{Sr}^{-}+V_{Sr}^{2-}$.

Supplemental Information

# Metal-organic Pulsed Laser Deposition for Complex Oxide Heterostructures

*Jung-Woo Lee*,[1,11] *Jieun Kim*,[1,11] *Anthony L. Edgeton*,[1,11] *Tula R. Paudel*,[2,10] *Neil Campbell*,[3] *Brenton A. Noesges*,[4] *Jonathon L. Schad*,[1] *Jiangfeng Yang*,[1] *Katelyn Wada*,[5] *Jonathan Moreno-Ramirez*,[5,6] *Nicholas Parker*,[5] *Yulin Gan*,[7] *Hyungwoo Lee*,[1] *Dennis V. Christensen*,[7] *Kitae Eom*,[1] *Jong-Hoon Kang*,[1] *Yunzhong Chen*,[7] *Thomas Tybell*,[8] *Nini Pryds*,[7] *Dmitri A. Tenne*,[5]
*Leonard J. Brillson*,[4,9] *Mark S. Rzchowski*,[3] *Evgeny Y. Tsymbal*,[2] and *Chang-Beom Eom*[1,12]∗

[1]Department of Materials Science and Engineering, University of Wisconsin-Madison, Madison, WI 53706, USA.
[2]Department of Physics and Astronomy, Nebraska Center for Materials and Nanoscience, University of Nebraska, Lincoln, NE 68588, USA.
[3]Department of Physics, University of Wisconsin-Madison, Madison, WI 53706, USA.
[4]Department of Physics, The Ohio State University, Columbus, OH 43210, USA.
[5]Department of Physics, Boise State University, Boise, ID 83725, USA.
[6]Riverstone International School, Boise, ID 83716, USA.
[7]Department of Energy Conversion and Storage, Technical University of Denmark (DTU), DK-2800 Kongens Lyngby, Denmark.
[8]Department of Electronic Systems, Norwegian University of Science and Technology, 7491 Trondheim, Norway.
[9]Department of Electrical and Computer Engineering, The Ohio State University, Columbus, OH 43210, USA.
[10]Present address: Department of Physics, South Dakota School of Mines and Technology, Rapid City, South Dakota 57701, USA.
[11]These authors contributed equally.
[12]Lead contact
*Correspondence: eom@engr.wisc.edu

## 1. Metal-organic pulsed laser deposition system (MOPLD)

A custom-built growth chamber, integrating pulsed laser deposition (PLD) with atomic layer control and ultra-high vacuum (UHV) chemical vapor deposition (CVD), is shown in Figure S1. Key features of this system include: metal-organic gas injectors that enable precise control of precursor gas flow; a calibrated conductance showerhead nozzle; and a pressure feedback system utilizing a motorized leak valve and capacitance manometer to achieve a stable mass flow rate with minimal transients.[1,2] To prevent precursor condensation, all components of the gas injector assembly are heated.

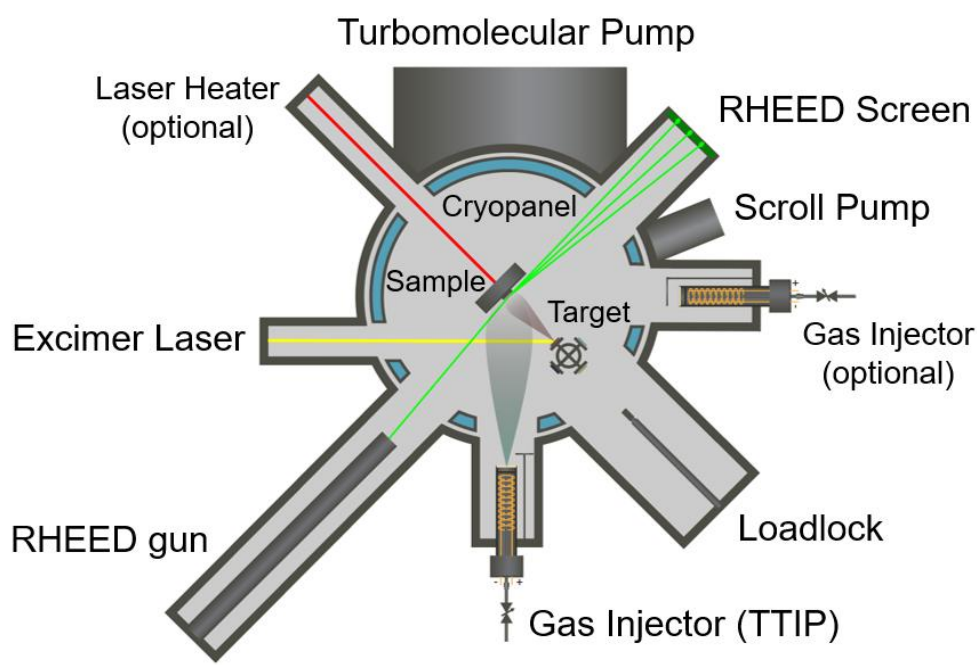

**Figure S1.** A schematic of the home-built metal-organic pulsed laser deposition system. The system integrates a PLD setup with high-pressure RHEED and combined gas supply systems. A cryopanel is installed to efficiently collect by-products generated from TTIP decomposition. Additionally, the system includes an optional laser heating module and a second gas injector, which were not utilized in this experiment.

## 2. Proposed growth mechanism of STO in the MOPLD system

It is known that desorption-limited growth of $TiO_2$ occurs at high temperatures, typically above ~700 °C, when TTIP is used as a precursor.[3] To control the adsorption/desorption process of TTIP on the substrate surface, two conditions should be satisfied: (i) the total working pressure must be low enough to minimize gas-phase reactions of TTIP, ensuring surface reactions dominate, and (ii) the growth temperature should be sufficiently high for TTIP to adsorb on an SrO-terminated STO surface while desorbing from a $TiO_2$-terminated STO surface. In this way, TTIP selectively decomposes only on SrO-terminated STO surfaces,[4] leading to stoichiometric STO growth.

The adsorption and desorption processes are thermally activated and can be described by an Arrhenius plot, where the slope of the line represents the activation energy for these processes (Figure S2). The region between the two lines defines the stoichiometric growth window for deposition.

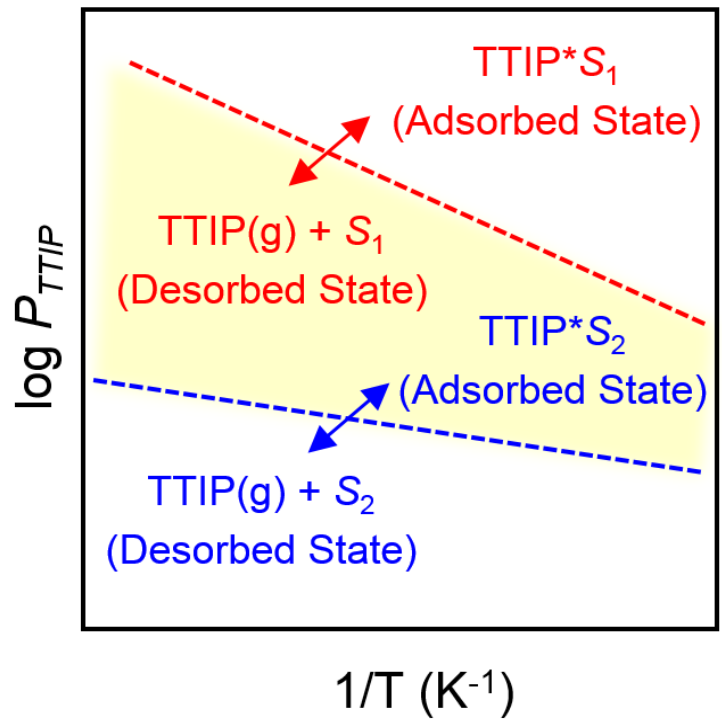

**Figure S2**. Schematic illustrating the thermodynamic consideration for selective TTIP adsorption in log $P_{TTIP}$ vs. $1/T$ ($K^{-1}$) diagram. $S_1$ and $S_2$ denote adsorption sites on $TiO_2$-terminated and SrO-terminated STO surfaces, respectively. Above the red and blue lines, TTIP adsorption is energetically favorable on $TiO_2$- and SrO-terminated surfaces, respectively. In the yellow region between two lines, selective adsorption of TTIP occurs on the SrO-terminated surface, enabling the growth of stoichiometric STO films.

## 3. Control of cation stoichiometry of STO within the solubility limit (<0.1 atomic %)

We propose that adjusting the TTIP gas inlet pressure influences the chemical potential of $TiO_2$ in the system, assuming that the total pressure changes are negligible within this range. While STO is known to be a line compound in the $SrO$–$TiO_2$ binary system,[5] a narrow solubility range exists near STO due to entropy of mixing.

The Gibbs-Duhem relationship for $\mu_{TiO_2}$ is given by:

$$\mu_{TiO_2} = -SdT + VdP \tag{S1}$$

In our experiment, the growth temperature was fixed at 900 °C. Under these conditions, $dG/dP > 0$, indicating that the chemical potential of $TiO_2$ decreases as the TTIP partial pressure decreases. The equilibrium Sr/Ti ratio in the STO films is therefore determined by the relative position of the STO free energy curve and the chemical potential of $TiO_2$ (Figure S3).

When the TTIP gas inlet pressure is too low, the tangential line intersects the free energy curve of $Sr_2TiO_4$, as observed for samples grown at 9 and 10 mTorr. Conversely, at higher TTIP pressures (e.g., 21 mTorr), the equilibrium composition shifts slightly to the Ti-rich side. Beyond this pressure, growth occurs under non-equilibrium conditions, leading to Ti-rich STO films with lattice expansion. (Figure 2A)

Based on previous studies,[6] the solubility limit between STO and $TiO_2$ at 1000 °C in air is less than 0.1 atomic %. Under our experimental conditions—900 °C and TTIP pressure of ~1 × 10⁻⁶ Torr—the solubility is expected to be even narrower. First, considering the temperature: 1000°C is believed be below the eutectic temperature of the STO–$TiO_2$ system.[5] In this regime, solubility decreases further as the temperature decreases. Second, regarding the oxygen partial pressure: lower oxygen partial pressure suppresses solid-solution formation, similar to the behavior observed in $LRE_{1+x}Ba_{2-x}CuO_{7-\delta}$ system (LRE: Nd, Sm, Gd).[7]

Therefore, we conclude that the solubility range between STO and $TiO_2$ under our experimental conditions remains limited to less than 0.1 atomic %.

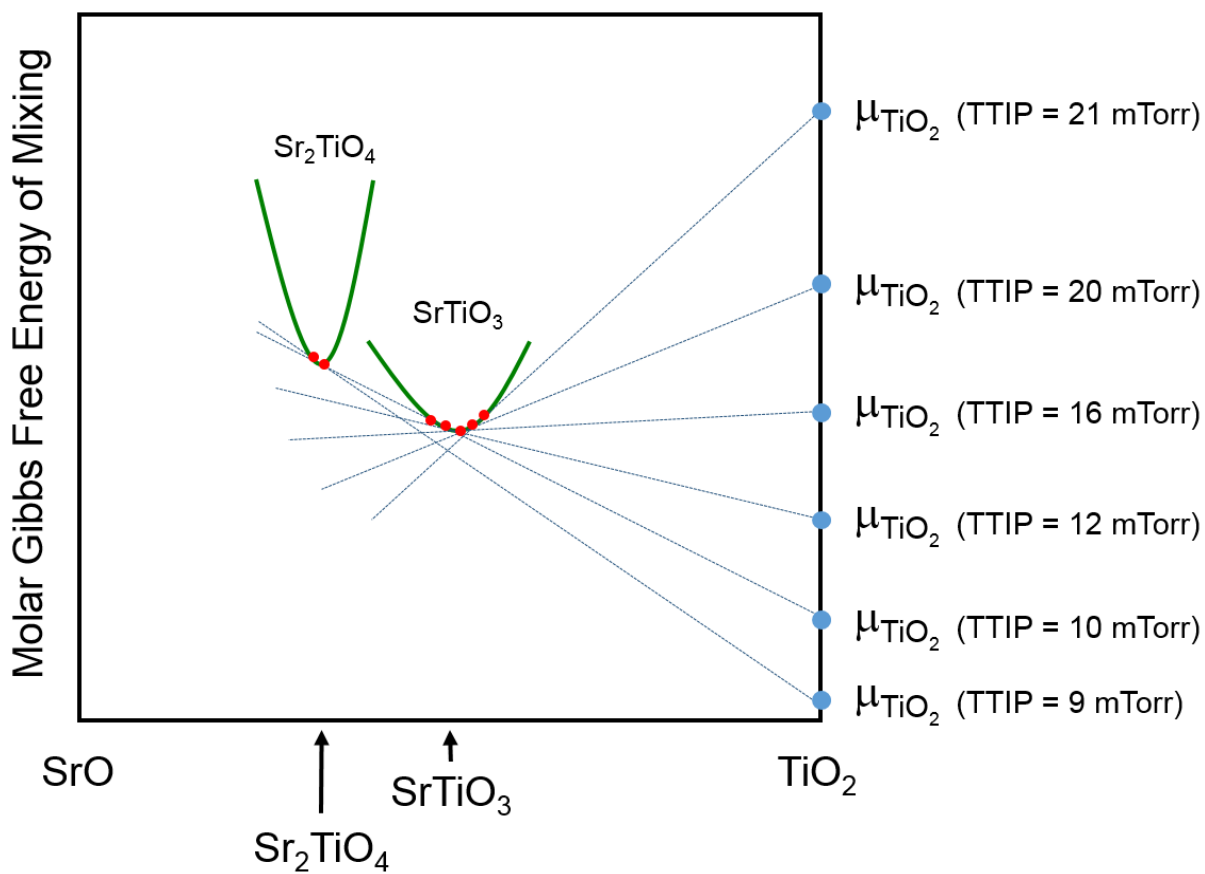

**Figure S3.** Schematic free energy diagram of SrO-TiO$_2$ binary system. The diagram illustrates how cation stoichiometry in STO can be controlled by adjusting the TTIP flux. For simplicity, changes in total pressure within this TTIP gas inlet pressure range are neglected. The curvature of the STO and Sr$_2$TiO$_4$ phases is exaggerated for clarity.

## 4. Control of point defect by adjusting cation stoichiometry

In Figure S4, we present the phase diagram of STO as a function of chemical potentials ($\Delta\mu$) of constituent elements. To construct this diagram, we solved a series of linear equations representing the formation of STO:

1. STO formation:

$$\Delta\mu_{Sr} + \Delta\mu_{Ti} + 3\Delta\mu_O \leq \Delta H_f \text{ (STO)} \tag{S2}$$

2. Competing phases:

- SrO (Sr-rich environment):

$$\Delta\mu_{Sr} + \Delta\mu_O \geq \Delta H_f \text{ (SrO)} \tag{S3}$$

- $Sr_2TiO_4$ (Sr-rich environment):

$$2\Delta\mu_{Sr} + \Delta\mu_{Ti} + 4\Delta\mu_O \geq \Delta H_f \text{ (Sr}_2\text{TiO}_4\text{)} \tag{S4}$$

- $TiO_2$ (Ti-rich environment):

$$\Delta\mu_{Ti} + 2\Delta\mu_O \geq \Delta H_f \text{ (TiO}_2\text{)} \tag{S5}$$

3. Elemental phases (stability constraints):

$$\Delta\mu_{Ti} \geq 0, \Delta\mu_{Sr} \geq 0, \Delta\mu_{Ti} \geq 0 \tag{S6}$$

Here $\Delta H_f$ represents experimental heat of formations. The green-shaded area in the diagram indicates the range of chemical potentials where STO is stable.

In our experiments, increasing the TTIP flux raises the chemical potential of both Ti and O. We assume that the chemical potential shift of STO is *toward* $TiO_2$ by increasing TTIP, since $TiO_2$ is the most stable form after decomposition of TTIP. Hence, both $\Delta\mu_{Ti}$ and $\Delta\mu_O$ are increased in a correlated manner.

To quantify this relationship, we considered the solid solution between STO and $TiO_2$. If the amount of $TiO_2$ increases by a small fraction *a* (where $a \ll 1$), the compound formula can be expressed as:

$$(1-a)Sr_{0.2}Ti_{0.2}O_{0.6} + a(Ti_{1/3}O_{2/3}). \tag{S7}$$

Here, $a = 0$ corresponds to stoichiometric STO. For equilibrium, the relationship $\Delta\mu_{Sr} + \Delta\mu_{Ti} + 3\Delta\mu_O = 0$ holds. Solving for the changes in chemical potential as $TiO_2$ content increases:

$$\Delta\mu_{Sr} = -0.2a \text{ (decreased)}, \Delta\mu_{Ti} = \frac{2}{15}a \text{ (increased)}, 3\Delta\mu_O = \frac{1}{15}a \text{ (increased)}.$$

Thus, the relationship between $\Delta\mu_{Ti}$ and $\Delta\mu_O$ is:

$$\Delta\mu_{Ti} = 6\Delta\mu_O. \tag{S8}$$

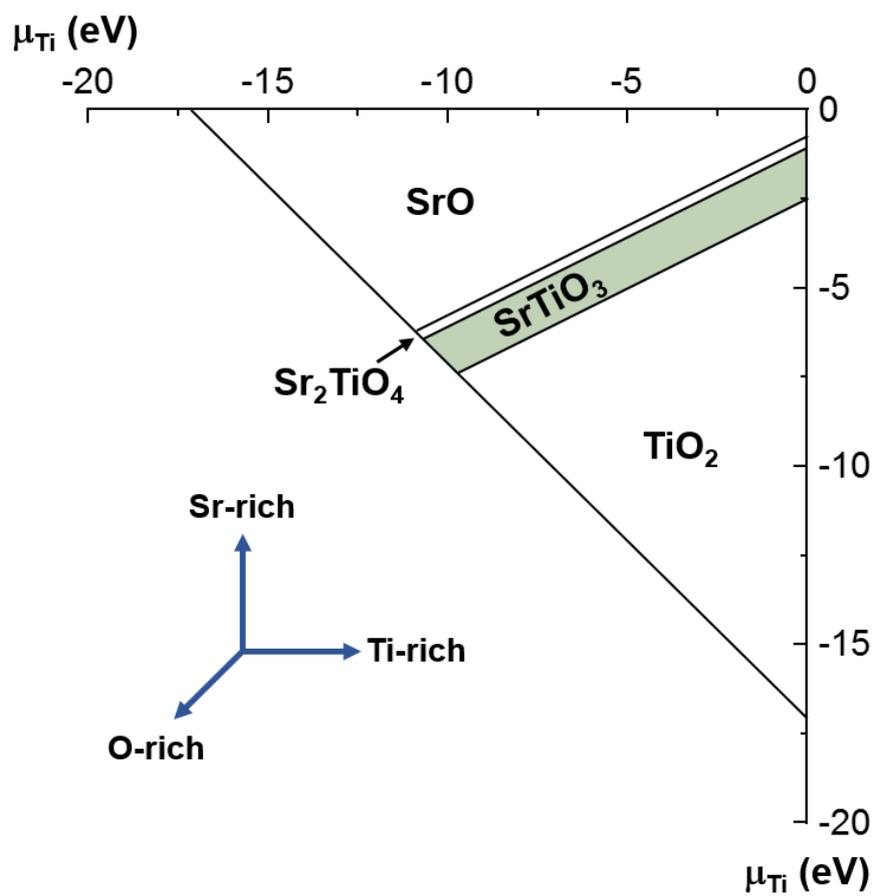

**Figure S4**. Phase diagram near the STO composition as a function of the chemical potential of Sr, Ti and O. The green-shaded area represents the stable region of STO.

## 5. STO Thin Film Growth by MOPLD

The atomic force microscopy (AFM) image of the STO film surface (inset of Figure S5) reveals a distinct step-terrace structure, confirming the layer-by-layer growth mode (Figure S5).

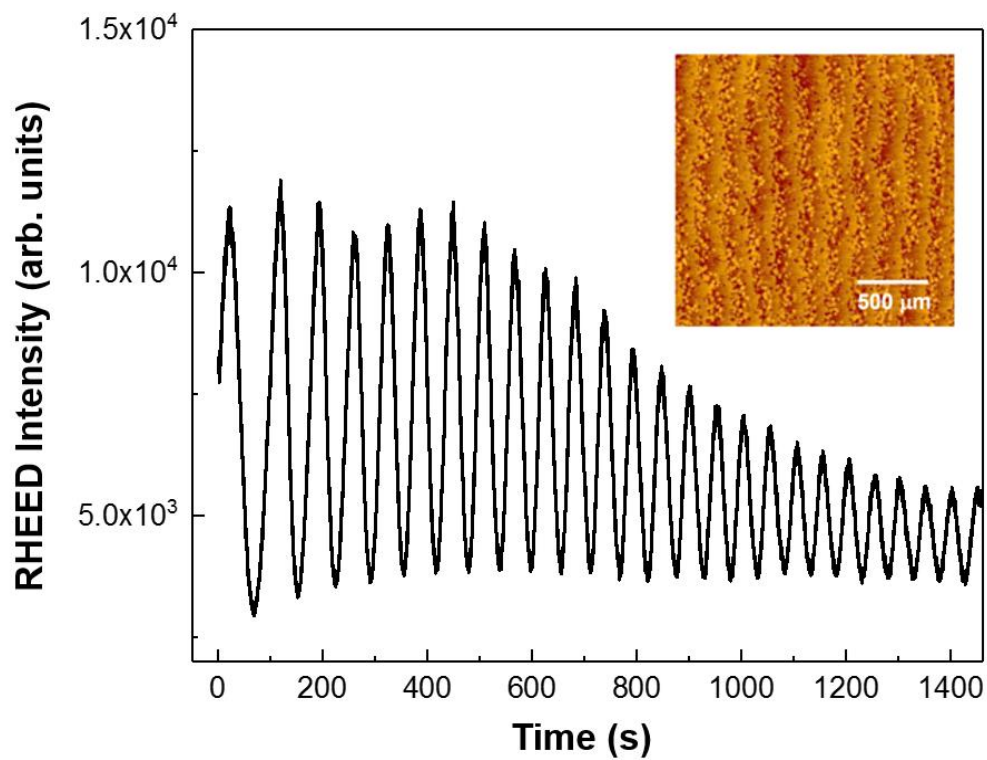

**Figure S5.** RHEED oscillation during the growth of a 25 u.c. STO film on STO (001) substrate by MOPLD. The inset shows an atomic force microscopy (AFM) image of the 25 u.c. STO film on the STO (001) substrate.

## 6. Measurement of the Shubnikov-de Hass (SdH) effect

In Figure S6, longitudinal resistance is evaluated as;

$$R_{xx\_a} = \frac{\pi}{\ln 2} \times \frac{V_{8,12}}{I_{7,11}}, \ R_{xx\_b} = \frac{\pi}{\ln 2} \times \frac{V_{11,12}}{I_{7,8}}, \ R_{xx\_c} = \frac{\pi}{\ln 2} \times \frac{V_{7,11}}{I_{8,12}}, \ R_{xx\_d} = \frac{\pi}{\ln 2} \times \frac{V_{7,8}}{I_{11,12}} \quad (S9)$$

Hall resistance is measured by;

$$R_{xy\_a} = \frac{\pi}{\ln 2} \times \frac{V_{11,8}}{I_{7,12}}, \ R_{xy\_b} = \frac{\pi}{\ln 2} \times \frac{V_{12,7}}{I_{11,8}} \quad (S10)$$

Because $R_{xx\_tot\_ab}$ can be expressed as;

$$R_{xx\_tot\_ab} = f \times \frac{(R_{xx\_a}+R_{xx\_b})}{2}, \text{ where } \cosh\left\{\frac{Q-1}{Q+1} \times \frac{\ln 2}{f}\right\} = \frac{1}{2}e^{\left(\frac{\ln 2}{f}\right)}, \ Q = \frac{R_{xx\_a}}{R_{xx\_b}} \quad (S11)$$

Therefore, $R_{xx\_tot}$ and $R_{xy\_tot}$ are calculated by;

$$R_{xx\_tot} = \frac{(R_{xx\_tot\_ab}+R_{xx\_tot\_cd})}{2}, \ R_{xy\_tot} = \frac{(R_{xy\_a}+R_{xy\_b})}{2} \quad (S12)$$

In two-dimensional system, the thickness factor $t = 1$ (unit length) is considered.

Using the following equation;

$$\frac{\Delta R_{xx}(T)}{\Delta R_{xx}(T_0)} = \frac{TR_0(T)}{T_0 R_0(T_0)} \frac{\sinh(\alpha T_0)}{\sinh(\alpha T)} \quad (S13)$$

The fitting result at fixed $B = 10.60$ T is shown in Figure S7, where $\alpha = 1.385 \pm 0.188$ (with 95 % confidence bounds), $m^* = (1.00 \pm 0.13)m_e$ (at $T_0 = 1.9$ K)

The fitting result of the equation 5 at fixed T = 1.9 K;

$$F = \frac{\hbar}{2\pi e}A \ (A = \pi k_F^2), \ n_{2D} = g_v g_s eF/h \quad (S14)$$

For a single valley and $g_S = 2$: $n_{2D} = 1.7 \times 10^{12} cm^{-2}$ (Figure S8).

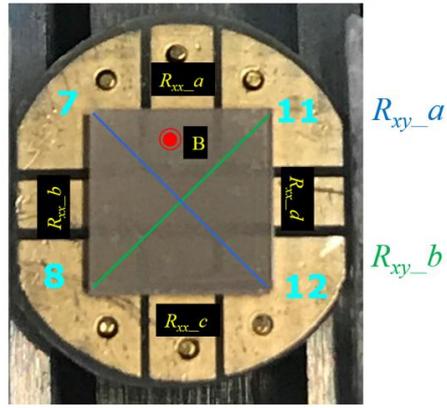

**Figure S6**. Van der Pauw configuration used for the measurement.

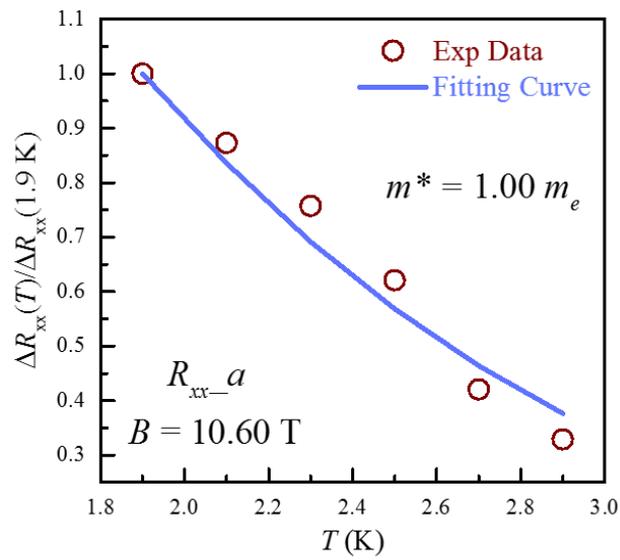

**Figure S7**. Temperature dependence of the scaled oscillation amplitude at $B = 10.60$ T. The effective electron mass is determined from the temperature dependence of the oscillations.

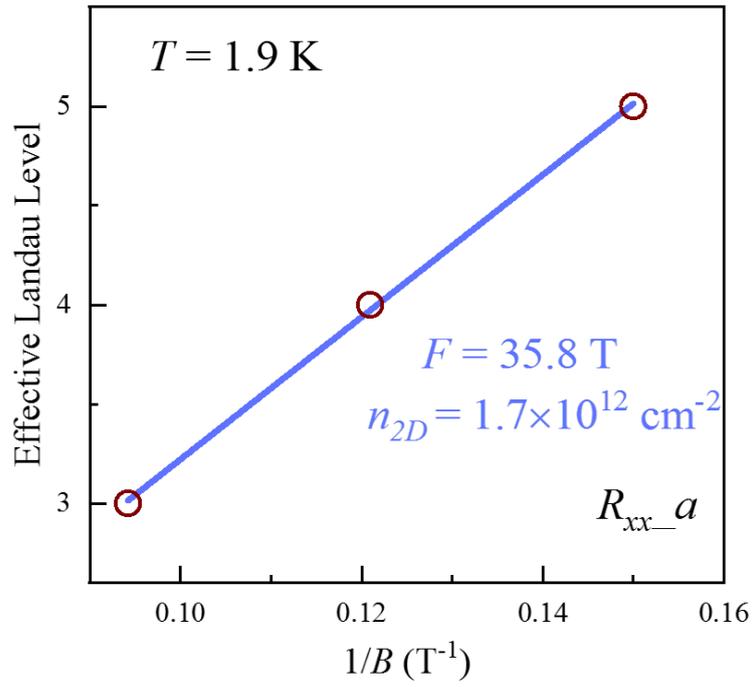

**Figure S8**. The SdH frequency (F) and quantum carrier density calculated at 1.9 K. The electron concentration determined by fitting the peak positions in the reciprocal magnetic field.

## 7. Depth-Resolved Cathodoluminescence Spectroscopy (DRCLS)

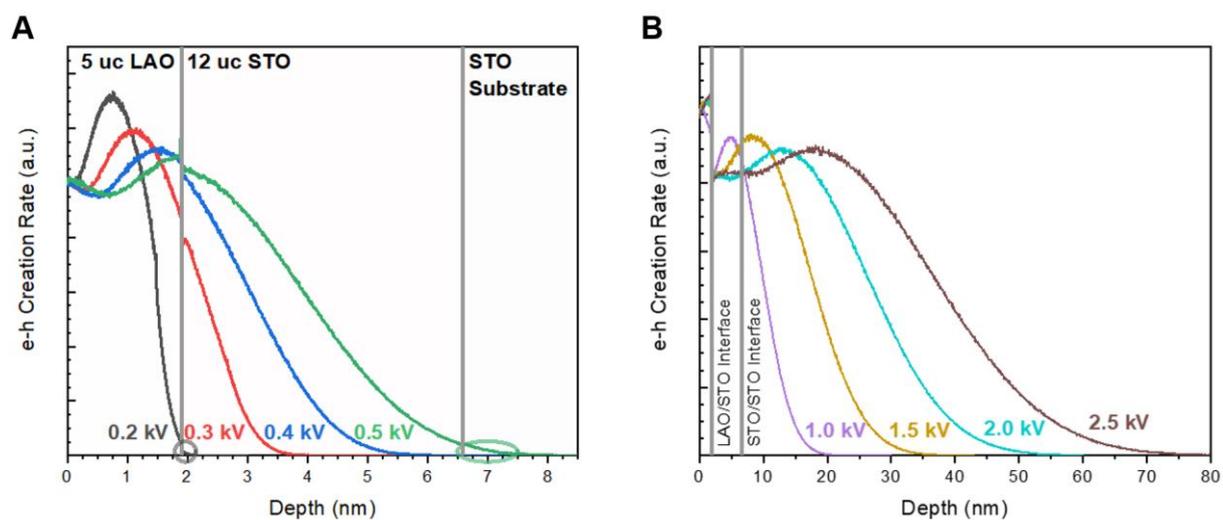

**Figure S9**. Electron-hole reaction rate as a function of incident beam voltage simulated by Monte Carlo simulations of electrons in solids (CASINO).

(A) Low-voltage simulations for the near surface region.

(B) High-voltage simulations for the bulk region.